\documentclass[nologo,11pt,a4paper,hidelinks]{ETHpaper}

\usepackage{verbatim}  
\usepackage{graphicx}                  
\usepackage[compress]{natbib}     

\usepackage{amsmath}
\usepackage{caption}
\usepackage{dsfont}

\graphicspath{{figures/}}

\usepackage{color}

\renewcommand{\epsilon}{\varepsilon} 
 
\renewcommand*{\=}{{\kern0.1em=\kern0.1em}}
\renewcommand*{\-}{{\kern0.1em-\kern0.1em}} 
\newcommand*{\+}{{\kern0.1em+\kern0.1em}}

\begin{document}
 

\title{Modeling the formation of R\&D alliances: \\ An agent-based model with empirical validation}
\titlealternative{Modeling the formation of R\&D alliances: An agent-based model with empirical validation}

  \author{Mario V. Tomasello,$^{1}$ Rebekka Burkholz,$^{1}$ 
    Frank Schweitzer$^{*,1}$}
\address{
$^\ast$Corresponding author; E-mail: fschweitzer@ethz.ch\\
$^1$Chair of Systems Design, ETH Zurich, Department of Management, Technology and \\ Economics, Weinbergstrasse 56/58, CH-8092 Zurich, Switzerland 
}

\authoralternative{M. V. Tomasello, R. Burkholz, 
  F. Schweitzer}
\www{\url{http://www.sg.ethz.ch}}

\maketitle

\begin{abstract}
  We develop an agent-based model to reproduce the size distribution of R\&D alliances of firms.
  Agents are uniformly selected to initiate an alliance and to invite collaboration partners.
  These decide about acceptance based on an individual threshold that is compared with  the utility expected from joining the current alliance.
  The benefit of alliances results from the fitness of the agents involved.
  Fitness is obtained from an empirical distribution of agent's activities.
  The cost of an alliance reflects its coordination effort. 
  Two free parameters $a_{c}$ and $a_{l}$ scale the costs and the individual threshold.
    If initiators receive $R$ rejections of invitations, the alliance formation stops and another initiator is selected.
    The three free parameters $(a_{c},a_{l},R)$ are calibrated against a large scale data set of about 15,000 firms engaging in about 15,000 R\&D alliances over 26 years.
    For the validation of the model we compare the empirical size distribution with the theoretical one, using confidence bands, to find a very good agreement.
    As an asset of our agent-based model, we provide an analytical solution that allows to reduce the simulation effort considerably.
    The analytical solution applies to general forms of the utility of alliances. Hence, the model can be extended to other cases of alliance formation.
    While no information about the initiators of an alliance is available, 
 our results indicate that mostly firms with high fitness are able to attract newcomers and to establish larger alliances. 
\end{abstract}




\section{Introduction}

Collaboration can be widely observed in different social and economic systems, where agents strive to reach a common goal. 
Scientists collaborate to write joint publications \citep{Katz1997}, firms collaborate to file joint patents \citep{Hoang2005,Kim2007}, and software developers collaborate to create joint software products \citep{Bitzer2010,Lakhani2003}.
To \emph{explain} collaboration, economic research has traditionally focused on different aspects of labor division \citep{durkheim2014division} and productivity of teams \citep{Scholtes2016}.
However, in the wake of technology-driven economic growth, the question how to \emph{boost} collaboration to 
foster knowledge transfer and innovation has become more important \citep{Frenz2009}.

The current research about the dynamics of R\&D networks can be seen as a major contribution to better understand how firms collaborate in patenting activities.
In this network representation, nodes depict the economic agents, i.e. the firms, and links between nodes their collaboration.
Specifically, firms formally declare this collaboration in publicly announced \emph{alliances}, which can involve more than two partners.
So, it makes sense to ask how the \emph{size of alliances}, i.e. the \emph{number of partners involved}, can be explained by means of an agent based model, which is the aim of the current paper.

To address this questions, we can build on a number of empirical studies about R\&D alliances.
It was shown that, because firms are involved in different alliances at the same time, their collaboration results in a large network component, in which even firms not directly collaborating are still connected through other firms (See Figure \ref{fig:network_plots}).
At the same time, a large number of small firm alliances exist that are not connected to the rest of the network.
These co-existing sub-networks are called \emph{components} in the following. 

The formation of a strongly connected component can be seen as an emergent property of the economic network because it is not planned top down, but emerges during the process of alliance formation, if (some) firms become engaged in more than one alliance.
Once such a strongly connected component exists, it greatly enhances the transfer of knowledge and the diffusion of innovations even between distant firms, so it is beneficial from a policy perspective.

\citep{tomasello2014therole,Schweitzer2017growth} have already proposed an agent-based model that is able to reproduce most of the properties of the observed R\&D network.
These properties include (i) the distribution of component sizes, i.e. the number of components of a given size plus the size of the largest connected component, (ii) the distribution of local clustering coefficients,
i.e. the fraction of firms in a component that form triads  (closed triangles) in their collaboration, (iii) the distribution of the lengths of shortest paths  that connect any two firms in the network, and (iv) the distribution of degrees, i.e. the number of partners of a firm.

This agent-based model, while successfully reproducing network features along different dimensions, takes two empirical distributions as an input: (a) the distribution of agent's \emph{activities}, i.e. their propensity to engage into a collaboration, and (b) the distribution of \emph{alliance sizes}, i.e. the number of partners involved in an alliance. 
The latter has been investigated empirically \citep[see][]{hagedoorn2002inter, tomasello2013riseandfall}.
Remarkably, one finds a broad and right-skewed distribution of alliance sizes (see Figure \ref{fig:m_partners}).
The same distribution was found even for different industrial sectors (including manufacturing, research, financial and service sectors), that  exhibit substantial differences otherwise. 

Previous modeling attempts in this field have, to the best of our knowledge, limited themselves only to general features of the R\&D network, such as the degree distribution or small world properties.
With the current paper, we want to move the agent-based modeling one important step forward, by \emph{explaining} the distribution of alliance sizes as an emergent feature of an underlying agent-based model instead of taking it from observations. 
This requires us to explicitly model \emph{how} agents form alliances, which implies to consider \emph{why} agents form alliances.
But given the empirical work on alliance sizes, we have some \emph{ground truth} to later judge the performance of our agent-based model in reproducing the distribution of alliance sizes.


\section{Empirical findings}
\label{sec:data}

\subsection{The network of R\&D alliances}
\label{sec:network-rd-alliances}

\paragraph{The dataset.}

We build our empirical R\&D network using the \textit{SDC Platinum} database,\footnote{\url{http://thomsonreuters.com/sdc-platinum/}} that reports approximately 672,000 publicly announced alliances in all countries, from 1984 to 2009, with a granularity of 1 day, between several kinds of economic actors (including manufacturing firms, investors, banks and universities) for which we commonly use the term ``firm'' in the following. 
We then select all the alliances characterized by the ``R\&D'' flag; after applying this filter, a total of $N=$14,829 alliances, connecting $n=$14,561 firms, are listed in the dataset.
An \emph{R\&D alliance} is defined as an declared partnership between two or more firms.
This can range from formal joint ventures to more informal research agreements, specifically aimed at research and development purposes.
Note that we do not have any information about the firm that initiated the alliance, nor about the sequence in which firms joined an alliance. 

The analysis of the data set, as well as all the network analyses and plots, are done by means of the R software for statistical computing.\footnote{\url{http://www.r-project.org/}}

\paragraph{Reconstructing the collaboration network.}

In the present study, we investigate the R\&D network aggregated over all years and all industrial sectors, which has to be reconstructed from the data set.
Firms are represented as nodes in the network and R\&D alliances as undirected links between nodes.
Isolate nodes, i.e. firms not taking part in any R\&D partnership, are simply excluded from our network representation.

\begin{figure}[htbp]
\begin{center}
\includegraphics[width=0.55\textwidth,trim=160 80 75 0]{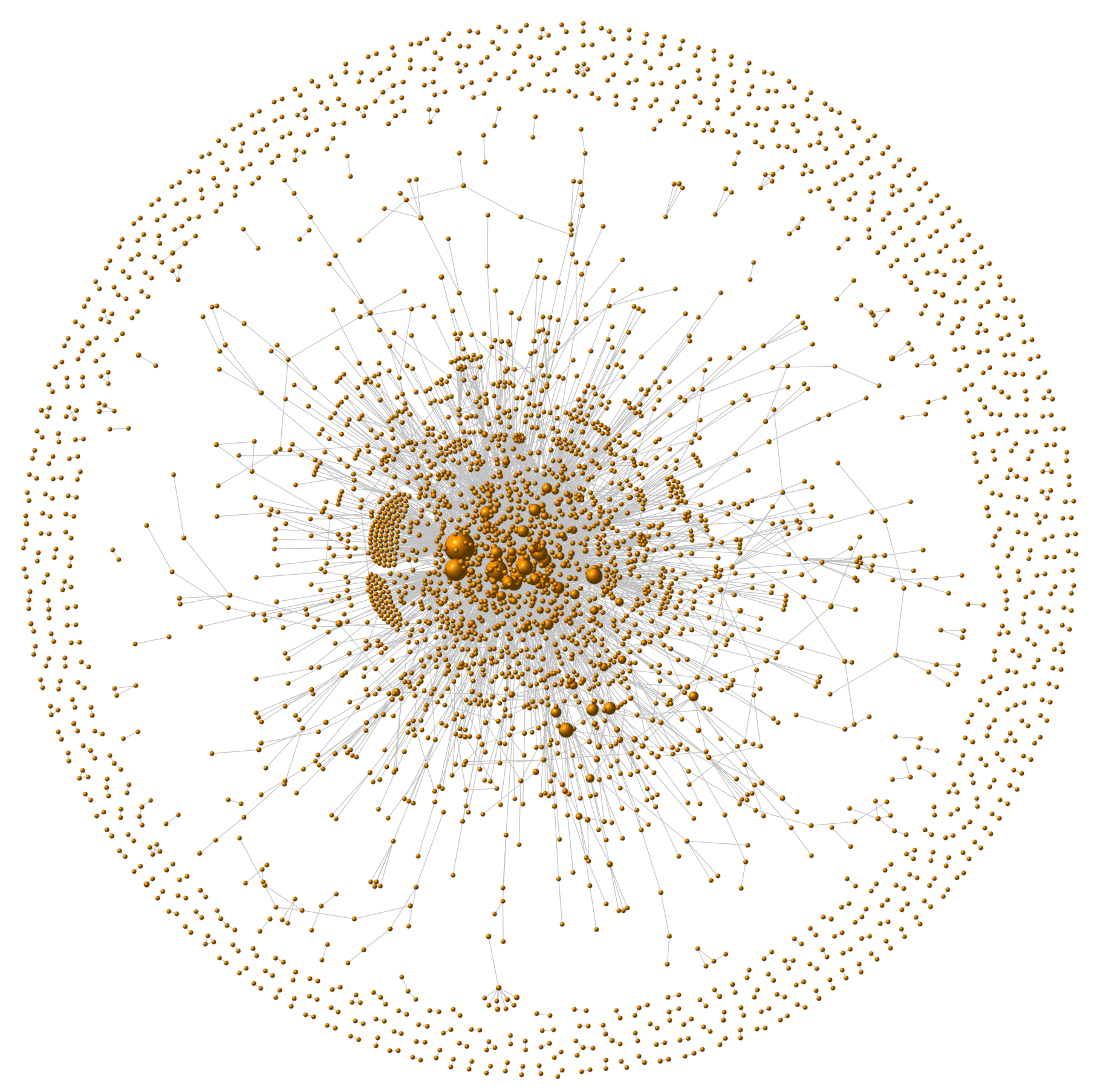}
\end{center}
\caption[]{Visual representation of the R\&D network that we analyze in this study -- the size of the nodes encodes their fitness.
We have used the \textit{igraph} package \citep{igraph_package} and the Fruchterman-Reingold layout algorithm \citep{fruchterman_reingold_1991}, which minimizes the number of crossing links.}
\label{fig:network_plots}
\end{figure}

When an R\&D alliance involves more than two firms, we assume that all the corresponding nodes are connected in pairs, forming a fully connected clique.
A ``standard'' two-partner alliance is then a fully connected clique of size 2. 
The choice of the fully connected clique -- rather than less interconnected network architectures -- derives from the fact that alliances of more than two partners, although representing only a minority, require great coordination and resources.
Therefore, they have to be associated with a higher number of links in the corresponding collaboration networks.
By following this procedure, the 14,829 R\&D alliances listed in the dataset result in a total of 21,572 links.
The resulting network is shown in Figure~\ref{fig:network_plots}.

\paragraph{Distribution of alliance sizes}

A salient feature of the R\&D alliances in the SDC dataset is the variable number of partners they involve. Most of the collaborations (93\%) are stipulated between two partners, the remaining ones 
involve three or more partners.
In the following, we denote by $s$ the size of the alliance, whereas $n$ indicates the number of firms and $N$ the number of alliances.

\begin{figure}[h!]
\begin{center} \footnotesize{
\includegraphics[width=0.47\textwidth]{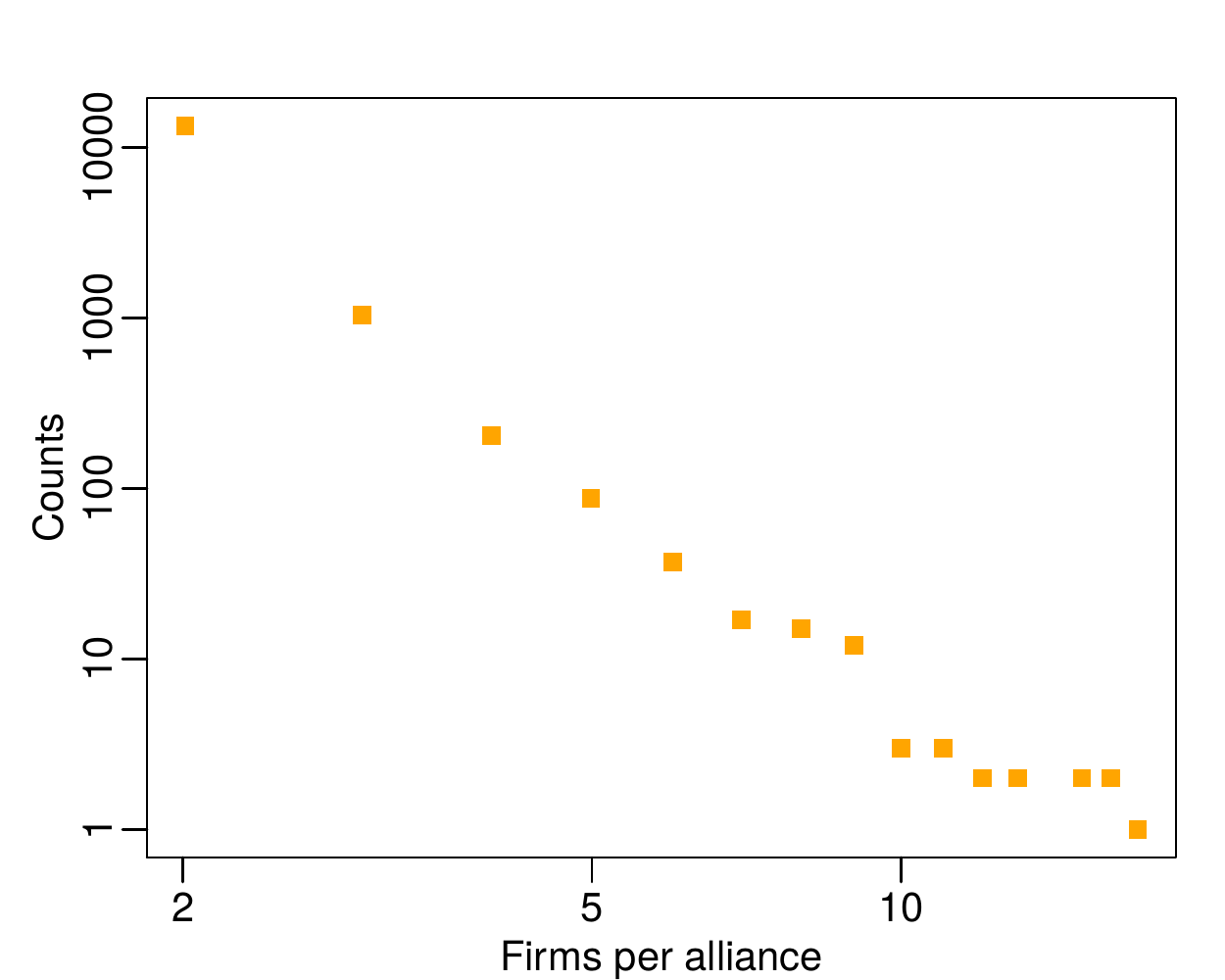}
}
\end{center}
\vspace{-8pt}
\caption{Histogram of  the  empirical alliance size distribution $p^{e}_{s}(s)$ measured on the R\&D network.
  This distribution is later used to evaluate the outcome of our agent-based model.
}
\label{fig:m_partners}
\end{figure}

We report the empirical distribution of the alliance size, $p^{e}_{s}(s)$ in the R\&D network in Figure \ref{fig:m_partners}.
As clearly visible, it spans one order of magnitude and is right-skewed.
It should be noted that an identification of the functional form of the distribution (e.g., power-law, exponential, log-normal and so on) is outside of the scope of this study. 
Our aim instead is to develop an agent-based model to reproduce this distribution, as described below.

\subsection{Defining a fitness measure for agents}
\label{sec:defin-fitn-meas}

Our agent-based model requires an attribute, called ``fitness'', that is assigned to each agent.
It describes how attractive an agent is for the other agents, to form an alliance.
To keep the model as a general as possible, we decide to proxy fitness by a measure which is \emph{not} system specific, such as the operational value of a firm. 
We choose the so-called agents' \textit{activity} \citep{perra2012activity}, which has been already successfully used on various data sets, such as online microblogging, actor networks, R\&D and co-authorship networks \citep{tomasello2014therole}.
The empirical \textit{activity} $\eta^{\Delta t}_{i,t}$ of an agent $i$ at time $t$, over a time window $\Delta t$, is defined as the \emph{number of alliances} $n^{\Delta t}_{i,t}$ that involve agent $i$ in the time window $\Delta t$ ending at time $t$, divided by the total number of alliances $N^{\Delta t}_{t}$ involving \textit{any} agent in the network during the same time period:

\begin{equation}
 \eta^{\Delta t}_{i,t} =  \frac{{n^{\Delta t}_{i,t}}}{ {N^{\Delta t}_{t}}}.
 \label{eq:2}
\end{equation}

It was found that activity distributions in most collaboration networks are right skewed and dispersed over several orders of magnitude, as in many other social and technological systems \citep{barabasi2005origin, pastor-satorras01:_dynam_correl_proper_inter}.
This is confirmed also for the case of R\&D networks, where the empirical activity values range from low 0.002 to the maximum value of $1$.
Applying this to the fitness of agents, this means that the agent with the highest fitness has a value of 2-3 orders of magnitude larger than the agents with the lowest fitness. %
Indeed, the vast majority of the agents has a fitness equal to the minimum value, which is also the \textit{median} value, and the average fitness is only slightly higher than that. Only one agent has a fitness equal to 1 (the highest possible value).

Contrary to most network indicators that display strong variability and dependence on time, especially in R\&D networks \citep[see][]{tomasello2013riseandfall}, activity is a stable attribute that can be assigned to firms to effectively estimate their propensity to engage in new collaborations, as well as their attractiveness to potential new collaborators.
Empirical activities are robust with respect to (a) the time $t$ at which they are measured, (b) the length of the selected time window $\Delta t$, (c) the sectoral classification of firms or authors, as shown by \citet{tomasello2014therole}.
Such a stability makes activity a perfect empirical proxy for our fitness attribute.

\begin{figure}[h!]
\begin{center} \footnotesize{
\includegraphics[width=0.47\textwidth]{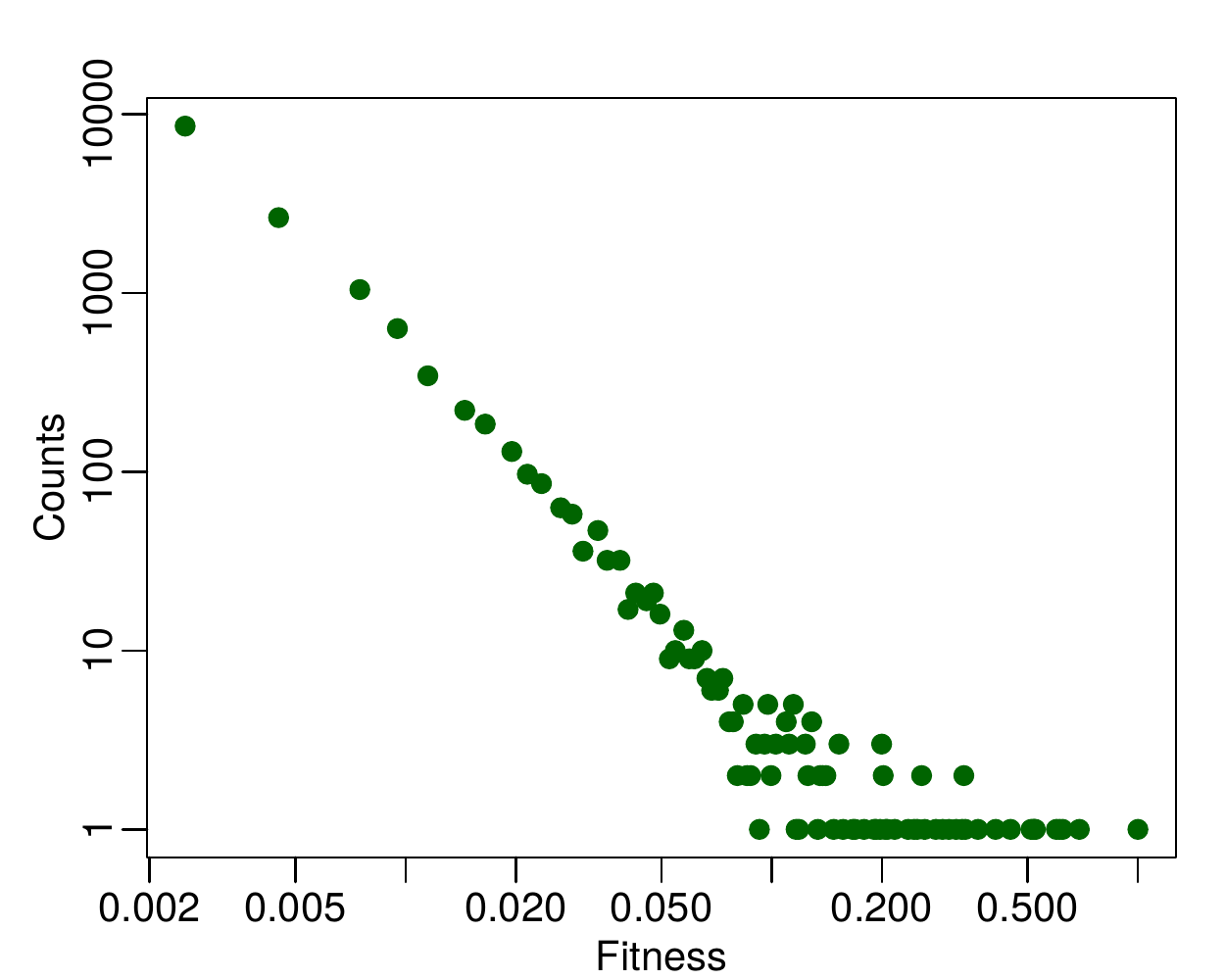}
}
\end{center}
\vspace{-10pt}
\caption[]{Histogram of agents' empirical fitness distribution, $p^{e}_{\eta}(\eta)$, measured on the R\&D network. This distribution is later used as an input for our agent-based computer simulations.}
\label{fig:empirical_fitness_attributes}
\end{figure}

Given the robustness with respect to the time window, we decide to compute the fitness values using the longest possible window, i.e.  the entire observation period, therefore $\eta_{i} \equiv \eta^{\Delta t = 26 \mathrm{years}}_{i,t=2009}$.
This considers the full information from the data set and results in activities $\eta_{i}$ that are always strictly greater than 0
because, by definition, all firms in our network must be involved in at least 1 alliance.
In Figure \ref{fig:empirical_fitness_attributes} we report the \emph{empirical distribution} of activity, i.e. of fitness, $p^{e}_{\eta}(\eta)$, for the analyzed R\&D network.
Further, in Figure \ref{fig:network_plots}, we have used the empirical fitness values of agents to scale their \emph{size} in the collaboration network.
Agents with higher fitness obviously form the core of the empirical R\&D network. 

\section{The modeling approach}\label{sec:model}

\subsection{Agent-based model of alliance formation}
\label{sec:agent-based-model}

In the following, we develop an agent-based model to reproduce the observed \emph{size distribution} of consortia, shown in Figure \ref{fig:m_partners}. 
This distribution is the result of a dynamic process in which agents decide to initiate or to join an alliance, i.e. it can only be understood by modeling the growth of the collaboration network.

\paragraph{Fitness of agents and initiation of an alliance.} 
  
Our model is a considerable extension of the network fitness model first proposed by \citet{bianconi01:_compet_and_multis_in_evolv_networ}.
Each agent $i$ is assigned a \textit{fitness} $\eta_{i}$ which is fixed and independent of time.
The values for the fitness are obtained from the empirical distribution $p^{e}_{\eta}(\eta)$, shown in Figure \ref{fig:empirical_fitness_attributes}. 

In our model, all $n$ agents can become active with a \emph{uniform probability}, which is chosen to be $1/n$, independent of their fitness. 
The sampling occurs with replacement, i.e. agents can also be chosen more than once to become active.
Activity means here that an agent  \emph{initiates} a new alliance; hence we refer to her as the ``initiator''.
We do not have empirical information about the agent that initiated an alliance, hence the assumption of a uniform probability for the activation is reasonable.

For our simulations, we choose a discrete time $t$ which measures the \emph{time to form an alliance}.
I.e. each time a new initiator is selected we start with $t=0$ and the maximum time for alliance formation is denoted as $T$.  
The newly created alliance can grow only if new collaborators join. 
This process is reflected in two steps: (i) the initiator \emph{invites} new collaboration partners, one per time step, (ii) the invitees \emph{accept} or \emph{reject} to join the alliance.

\paragraph{Utility of consortia.}
A number of agents form an alliance $\mathcal{C}(s^{t})$ of size $s^{t}$ which can change over time as new agents join the alliance.
There can be many consortia of different sizes coexisting over time. 
The utility function, $u^{t}$, of the alliance combines the benefits, $b^{t}$, and the costs, $c^{t}$, of the collaboration of the $s^{t}$ agents, i.e. both $b^{t}(s^{t})$ and $c^{t}(s^{t})$ depend on the current alliance size, $s^{t}$.
Further, the benefits should be a monotonous function of the fitness values of the currently involved agents, i.e. $b^{t}(...,\eta_{i},\eta_{j},...)$, whereas the costs should reflect the coordination effort of the alliance and thus should be a monotonous function of the \emph{size} of the alliance. 
For simplicity, we assume linear dependencies for the monotonous functions, i.e. the utility of an alliance is defined as
\begin{align}
  \label{eq:1}
  u^{t} =  b^{t}-c^{t}\;;\quad b^{t}=\sum_{m=1}^{s^{t}} {\eta_{m}}\;;\quad c^{t}= a_{c} \cdot \left[s^{t}-1\right]
\end{align}
where the parameter $a_{c}$ allows to scale costs against benefits. 
We note that the costs scale with the number of alliance \emph{partners} rather than with the number of their possible \emph{connections}, which would be quadratic, i.e. $s^{t}(s^{t}-1)/2$, if an alliance is seen as a fully connected clique.
The latter would account for a \emph{superlinear} increase in the \emph{coordination} effort between partners which sets strong limitations to larger consortia.
Here, instead we assume some sort of \emph{administration} cost based on the number of parties. 



\paragraph{Invitation of alliance partners.}

As the alliance grows, its utility will change in a \emph{non-monotonous} manner.
Precisely, according to Equation \ref{eq:1}, the utility will grow \emph{only} if the fitness $\eta_{j}$ of the new alliance member $j$ is larger than the scaling constant $a_{c}$. 
To ensure that this condition is met, an initiator preferably invites agents with a high fitness.
Precisely, similar to the fitness model of \citet{bianconi01:_compet_and_multis_in_evolv_networ},
the initiator $i$ chooses \emph{potential} alliance partners $j$, one at a time, with a probability proportional to their fitness $\eta_j$.

Different from the mentioned model, it is however left to the agents to decide whether they want to accept this invitation, i.e. to join the alliance. 
The initiator repeats the invitation procedure until  a number $R$ of invited partners \emph{refuse} the invitation to  the alliance.
I.e. $R$ is a parameter of our agent-based model.
If the \emph{current number of rejections}, $r^{t}$, reaches $R$, we assume that the alliance is fully formed and stops to grow in size.
At the same time the initiator loses its ``active'' status.
If the initiator receives $R$ rejections already from the first $R$ selected partners, then no alliance is formed.


\paragraph{Formation of collaboration links}

The second step in the formation of the alliance is the decision of the invitee to accept, or to not accept, the invitation.
An agent $j$ decides to join an alliance $\mathcal{C}$ at time $t+1$ if the utility of the alliance, $u^{t}$ is larger than a certain threshold, $u^{\mathrm{thr}}_{j}$, which is assumed to be \emph{heterogeneous} across agents.

Specifically, we argue that the threshold of an agent to join an alliance \emph{increases} with her fitness.
The rationale behind this is that agents with a high fitness are very attractive for initiators of consortia and thus receive invitations very often.
On the other hand, because of scarce resources, agents cannot simply accept all invitations, they have to be selective.
Therefore, the higher the own fitness and the attractiveness for consortia, the higher the threshold to accept an invitation.
Conversely, agents with a low fitness are not invited very often for an alliance, therefore they will be more inclined to accept invitations, i.e. their threshold is lower because of the lower fitness.
Hence, it is reasonable to argue that $u_{j}^{\mathrm{thr}}=a_{l}\,\eta_{j}$, i.e. $u_{j}^{\mathrm{thr}}$ is simply proportional to the fitness, where $a_{l}$ is a parameter of the model, to be determined later.

This results in the following condition for agent $j$ to join the alliance $\mathcal{C}$ in the next time step 
\begin{equation} \label{eq:consortium_fitness_acceptance}
 j \in \mathcal{C}_{t+1}  \quad \mathrm{if} \quad
 u^{t} \ge u_{j}^{\mathrm{thr}} \; \Rightarrow \sum_{m=1}^{s^{t}} \eta_{m} - a_{c}[s^{t}-1] \geq a_{l} \cdot \eta_{j}
\end{equation}

\paragraph{Implications.}
Our agent-based model builds on an interesting tension between the \emph{attractiveness} and the \emph{willingness} to become an alliance member, which is a novel point in the discussion of fitness models.
Hence, in our model the link formation process does \emph{not} follow a simple preferential attachment rule.
First, because agents do not \emph{increase} their individual fitness by accepting an invitation.
Secondly, because agents become the more selective, the fitter they are. 

It should be noted that, even though low-fit agents are more common in the network, high-fit nodes agents a much higher chance to be selected as potential partners and to establish new consortia. 
It is also clear from the setup of the model that agents with a low fitness would not be able to establish a larger alliance. 
They can likely not attract agents with high fitness, nor can they overcome the costs inclined in the formation of an alliance.
Hence, larger consortia depend on the initiation by agents with a high fitness and their ability to attract other agents with high fitness.

Eventually, our agent-based model combines probabilistic elements, such as the activation of agents and the invitation of partners, with deterministic elements, such as the decision of agents to join the alliance.
This decision differs from a ``best response'' rule, because agents do not decide based on complete (or global) information about all existing consortia.
Instead, they base their decision only on the (local) information of the current offer. 

Once the formation of an alliance is finished, this alliance is added to the existing network as a clique, i.e. a fully connected cluster of size $s^{T}$, which is in line with our procedure to reconstruct the network from empirical observations (see Section \ref{sec:data}).
This again differs from the mentioned fitness model in that the network grows with the sequential addition of cliques, and not of single links. {The addition of a single edge linking two nodes, as argued in Section \ref{sec:data}, can be thought of as the addition of a fully connected clique of size 2.}

\subsection{Analytic description of the alliance size distribution}
\label{sec:analyt-descr-coll}

We now proceed in two different directions.
First, we run stochastic simulations by implementing the agent-based rules described above.
Hence, at each time step we choose an agent to initiate an alliance.
Dependent on her success or failure, we add new cliques to the collaboration network and continue with choosing a new agent in the next time step.
This procedure is followed to obtain the results discussed in the subsequent sections.

However, we also formalize the model in a more analytic way to obtain an expression for the distribution of consortia sizes, $p_{s}(s)$.
We start from the fitness distribution $p_{\eta}(\eta)$, which we take as given from data.
In accordance with the empirical distribution $p^{e}_{\eta}(\eta)$, we consider $\eta$ as \emph{discrete} because of the binning given by the observations. 
We assume that $p_{\eta}(\eta)$ does not change during the formation of consortia.
That means even if agents with a given $\eta_{i}$ have accepted an invitation and can thus not be invited again to join the \emph{same} alliance, we assume that the distribution $p_{\eta}(\eta)$ of the \emph{remaining agents} is as before.
This assumption is justified if it is unlikely that an alliance grows to a significant proportion of the whole system, as it is the case in our studies.

With this, we derive an analytic proxy for the consortia size distribution $p_{s}(s)$.
It saves considerable computational effort and allows better insights into the model evolution. 
The distribution $p_{s}(s)$ gives us the \emph{probability} to find an alliance of \emph{final size} $s$.
This formation happened during the time steps $t = 0, \cdots, T$. 
At $t=0$, an initiator is picked at random and thus has a fitness $\eta_i$ with probability $p_{\eta}(\eta_i)$.
The alliance size at that time is $s^{t}\equiv s^{0}= 1$.
In each following time step, another agent, which has the fitness $\eta$ with probability $p_{\eta}(\eta)$, is invited to join the alliance. She either accepts so that the alliance size increases, $s^{t+1}= s^{t}+ 1$.
Or she rejects which leads to an increase in the number of rejections, $r^{t+1} = r^{t}+ 1$.
Hence, for the next time step $t+1$ in the alliance formation process always $t+1=s^{t}+r^{t}$ holds. 

To evaluate the probability for both cases, we have to keep track of the alliance utility.
For this, we introduce a time-dependent distribution $p(t,s^{t},r^{t},b^{t})$.
It represents the joint probability that at time step $t$ an alliance has reached the size $s^{t}$, while offers were  rejected $r^{t}$ times.

$b^{t}$ is the benefit of the alliance, i.e. the sum of the fitness values $\eta$ of the alliance partners according to Equation~\eqref{eq:1}.
The initial conditions for the time-dependent distribution are $p(0, 1, 0, b) =p^{e}_{\eta}(\eta_{i})$ because the alliance consists only of the initiator, and $p(0, s, r, b) = 0$ otherwise.

The arguments of $p(t,s^{t},r^{t},b^{t})$ can only have the following values: the size ranges from $s^{t} \in \{1,\ldots,n\}$ with $n$ as the total number of agents, the number of rejections from $r^{t} \in \{0,\ldots,R\}$ where $R$ is the maximum number of rejections that are still tolerated, and for the time $t \in \{0,\ldots,T\}$ where $T = n+R$ is the maximum time in which the formation of a single alliance is possible.
The benefit is bound to $b^{t}=\sum^{s^{t}}_{m=1} \eta_m \in \left[0,\sum^{n}_{m=1} \eta_m\right]$.
Furthermore, only combinations satisfying the condition $t+1 = s^{t}+r^{t}$ are reasonable, because, in each time step, either $s^{t}$ or $r^{t}$ are incremented. 
Otherwise, the growth of the alliance stops.
Hence, $p(t, s^{t}, r^{t}, b^{t}) = 0$ can be set for unreasonable combinations of $t$, $s^{t}$, and $r^{t}$. 

Following these considerations, we start from the initial conditions given above and iteratively deduce $p(t+1, s^{t+1}, r^{t+1}, b^{t+1})$ from $p(t,s^{t},r^{t},b^{t})$ at the previous time step.
For this, we have to take three different constellations into account:
(1) the alliance \emph{stops} growing at time $t$ because the maximum number of rejections $R$ was reached, 
(2) the alliance could \emph{potentially grow}, however the invited agent \emph{rejects} the offer, which leads to $s^{t+1}=s^{t}$ and $r^{t+1}=r^{t}+1$, and
(3) the alliancea \emph{actually grows} because the invited agent \emph{accepts} the offer, which leads to $s^{t+1}=s^{t}+1$ and $r^{t+1}=r^{t}$.  

For each of these three constellations we have to express the probability of its occurrence and the fact that the boundary conditions for the given arguments $t$, $s$, $r$ and $b$ are met.
For the latter, we use the \emph{indicator function}, which can be either zero or one.
$\mathds{1}_{[x,y]}(z)=1$ means that the value of $z$ is within the range of $x$ and $y$ and $\mathds{1}_{[x,y]}(z)=0$ otherwise, whereas
$\mathds{1}_{[x]}(z)=1$ means that the value of $z$ is precisely the value of $x$ and $\mathds{1}_{[x]}(z)=0$  otherwise.
This is more convenient and more compact than \texttt{if} and \texttt{else} to express different cases. 
We need two indicator functions  because we have constraints on two variables, $t$ and $r^{t}$, which determine $s^{t}=t-r^{t}$. 
With this, we can capture the three different constellations as follows:

\begin{align}
\begin{split}
  p(t+1, s^{t+1}, r^{t+1}, b^{t+1})  =  \mathds{1}_{[s^{t+1}+r^{t+1},T]}(t+1) \, \mathds{1}_{[R]}(r^{t+1})\, p(t,s^{t+1},R,b^{t+1})\\ 
  + \mathds{1}_{[t+1-s]}(r^{t+1})\, \mathds{1}_{[1,R]}(r^{t+1}) \, p(t,s^{t+1},r^{t+1}-1,b^{t+1}) \left[1 - F_{p_{\eta}}\left(\frac{b^{t+1}-a_c(s^{t+1}-1)}{a_l}\right) \right] \\
  + \mathds{1}_{[0,R-1]}(r^{t+1})\, 
  \sum^{\eta^{\star}}_{\eta= 0} p_{\eta}(\eta)\, p(t,s^{t+1}-1,r^{t+1},b^{t+1}-\eta)
\end{split}
    \label{eq:3}
\end{align}

The product $ \mathds{1}_{[s^{t+1}+r^{t+1},T]}(t+1) \,\mathds{1}_{[R]}(r^{t+1})$ in the first line of Eq.~\eqref{eq:3} refers to the case that too many rejections have happened already and the alliance stopped growing.
Hence, $b^{t+1}=b^{t}$, $s^{t+1}=s^{t}$ and $r^{t+1}=r^{t}=R$.
With this, $t+1= s^{t}+r^{t}$, but still within the maximum time allowed for alliance growth, $T=n+R$. 
$p(t,s^{t+1},R,b^{t+1})$, on the other hand, gives the probability of such a constellation at time $t$.

The second line of Eq.~\eqref{eq:3} counts all cases in which an invited agent rejects to join the alliance because its fitness is too high in comparison with the alliance, $a_l \eta > b^{t}-a_c(s^{t}-1)$, see Eq. \eqref{eq:consortium_fitness_acceptance}. 
The probability that this happens is given by the complementary cumulative distribution function, $1-F_{p_{\eta}}(\eta^{\star})$, with

\begin{align}
F_{p_{\eta}}(\eta^{\star})= \sum_{\eta\leq \eta^{\star}} p_{\eta}(\eta) \; ;\quad \eta^{\star}= \left[b^{t}-a_c(s^{t}-1)\right]/a_{l}
  \label{eq:9}
\end{align}

Because the rejection happens at time $t$, the current rejection value is $r^{t}=r^{t+1}-1$, while the size $s^{t+1}=s^{t}$ and the alliance benefit $b^{t+1}=b^{t}$ stay constant. 
The indicator function $\mathds{1}_{[1,R]}(r^{t+1})$ ensures that $r^{t+1}$ is still in the possible range of 1 (if that was the first rejection) and $R$. Otherwise, this would have been captured in constellation (1).
The second indicator function $\mathds{1}_{[t+1-s]}(r^{t+1})$ just reflects the boundary condition $t+1=s^{t+1}+r^{t+1}$.
Because the agent has rejected the offer, we have $s^{t+1}=s^{t}$.

The last line of Eq.~\eqref{eq:3} eventually considers all cases in which an agent accepts to join the alliance at time $t+1$.
In this case, $s^{t+1}=s^{t}+1$, $r^{t+1}=r^{t}$ and $b^{t+1}=b^{t}+\eta$.
The probability of this occurrence has to be multiplied by the probability $p_{\eta}(\eta)$ to find agents of fitness $\eta$.
The summation goes over all possible benefit values, $\eta$, for which agents accept to join the alliance, which follow from  Eq. \eqref{eq:consortium_fitness_acceptance}. 
This condition defines the value $\eta^{\star}$ introduced above.
I.e., agents join the alliance if their fitness $\eta$ is between $[0,\eta^{\star}]$. 
We can express this condition with respect to $t+1$ instead of $t$, i.e. $\eta^{\star}=\left\{b^{t+1}-a_c(s^{t+1}-2)\right\}/(1+a_{l})]$.

The indicator function $\mathds{1}_{[0,R]}(r^{t+1})$ eventually makes sure that the acceptable number of rejections $R$ is not already exceeded, i.e. the agent can still join the alliance.
Otherwise, it had been considered in constellation (1).
We note that our approach also applies to more general forms of cost functions $c^{t}(s^{t})$ than just $a_c (s^{t}-1)$.
Only the bound $\eta^{\star}$ would need to be adjusted correspondingly.

The iterations of Equation \eqref{eq:3} occur until $t=T=n+R$ is reached, which is the maximum time possible for forming an alliance.
This ensures that all consortia formations are counted in.
The final alliance size distribution is then simply the marginal distribution  

\begin{align}
  p_s(s) = \sum_{b^{T}}  p(t,s^{T},R,b^{T})
  \label{eq:4}
\end{align}

This distribution depends, for a given number of agents, on the three parameters $a_c$, $a_l$, $R$
and further on the empirical fitness distribution $p_{\eta}^{e}(\eta)$, which is given.
Hence, $a_c$, $a_l$, $R$ have to be determined in the following.

We emphasize that with our analytical solution  we follow a probabilistic approach.
That means, we take \emph{all possible} constellation into account.
This is equivalent to running a large number of computer simulations and averaging the results, at the end.



\section{Results}
\label{sec:results}

\subsection{Calibration of the agent-based model}
\label{sec:calibr-agent-based}

Our first task is to calibrate the agent-based model introduced above.
We take as model input the fitness distribution, $p_{\eta}(\eta)$, which is proxied by the empirical distribution $p_{\eta}^{e}(\eta)$ shown in Figure \ref{fig:empirical_fitness_attributes}.
It then remains to determine the set of parameters of the model, $a_{c}$ to scale the costs of the consortium, Equation \eqref{eq:1}, $a_{l}$ to scale the individual threshold for accepting invitations, Equation \eqref{eq:consortium_fitness_acceptance}, and $R$ which is the maximum number of rejections an initiator receives to stop the formation of an alliance. 
In order to determine these parameter values, we use  a standard \emph{maximum likelihood approach}.
This can be based either on computer simulations of the agent-based model or on the numerical solution of the analytic expressions given in Equation \eqref{eq:3}. 
Both lead to the same results for all analyzed parameter combinations.
Thus, we report the results from the numerics which can be obtained much faster.  

%

We are interested in the distribution of the alliance size, $p_{s}(s)$, given the set of parameters $(a_{c},a_{l},R)$. 
In a first step, we have to re-normalize this distribution to alliance sizes $\tilde{s}\geq 2$ because we have no observations about $p_{s}^{e}(1)$.
This renormalized distribution reads


\begin{align}
  \label{eq:5}
 p_{\tilde{s}}(\tilde{s} | a_c,a_l,R) = \frac{p_{s}(\tilde{s})}{\sum^{n}_{i=2} p_{s}({i})}
\end{align}

with $p_{\tilde{s}}(1| a_c,a_l,R) = 0$.  
%
The likelihood $\mathcal{L}(a_c,a_l,R)$ of each parameter combination $(a_{c},a_l,R)$ is then determined by the probability
to observe our data which consists of $N$ alliances with sizes $s_{1},\ldots,s_{N}$, given these parameters: 

\begin{align}
  \label{eq:6}
 \mathcal{L}(a_{c},a_{l},R) = \prod^N_{i=1} p_{\tilde{s}}(s_{i} | a_{c},a_{l},R ) 
\end{align}

For our simulations, as well as for the numerical solution of Equation \eqref{eq:3}, we will choose the set of parameters  $(\hat{a}_c,\hat{a}_l,\hat{R})$  that maximizes this likelihood:

\begin{align}
  \label{eq:7}
 (\hat{a}_c,\hat{a}_l,\hat{R}) =  \arg\max \mathcal{L}(a_{c},a_{l},R) = \arg\max \mathcal{L}(a_{c},a_{l},R)^{1/N}
 \end{align}

\begin{figure}[htbp]
\begin{center}
\includegraphics[width=0.49\textwidth]{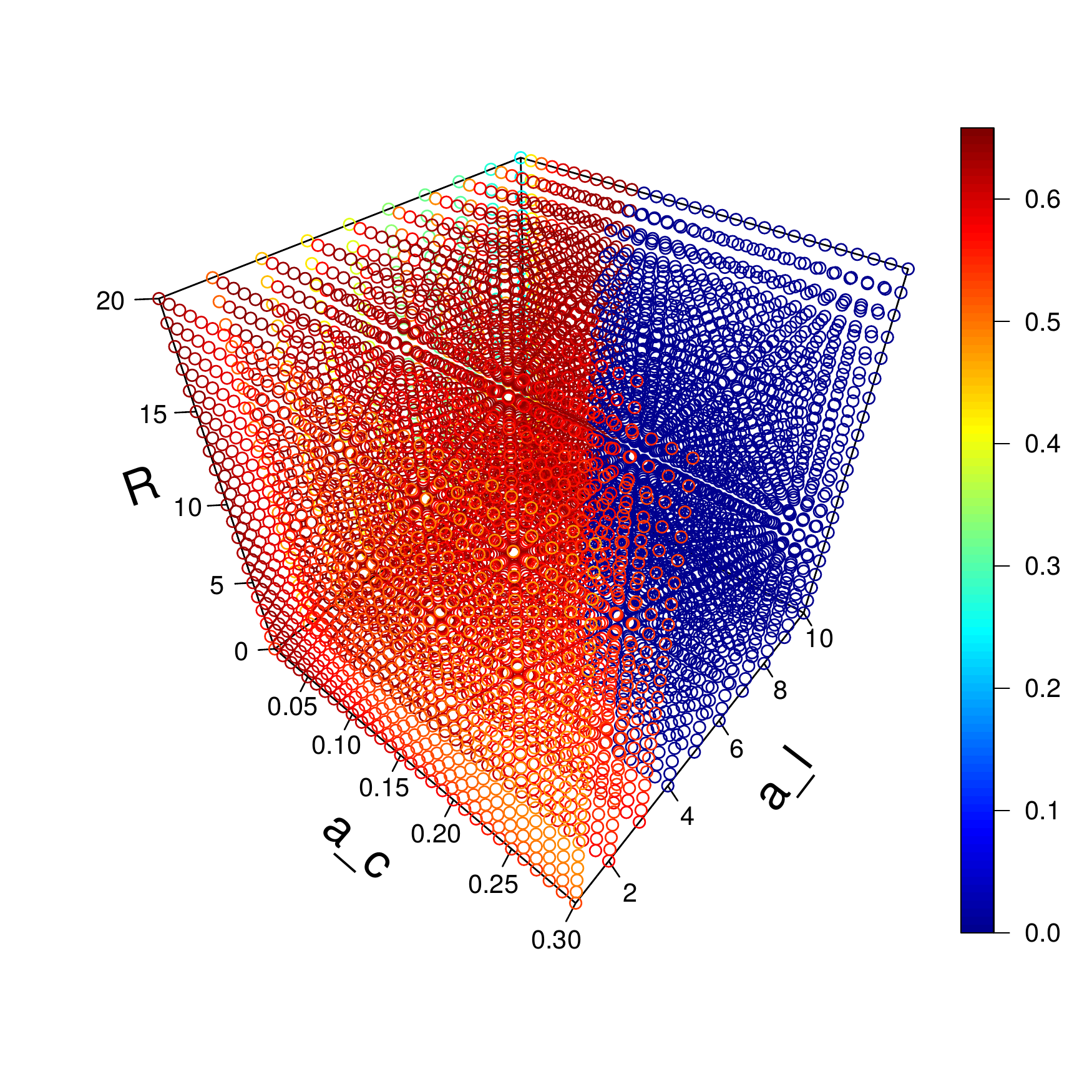}
\end{center}
\vspace{-18pt}
\caption[]{Results of the parameter values $(a_{s},a_{l},R)$  using the  maximum likelihood estimation, $\mathcal{L}(a_{c},a_{l},R)^{1/N}$.}
\label{fig:optimal_configurations_RSL}
\end{figure}

The exponent $1/N$ avoids the comparison of too small values and guaranties thus numerical stability.
The results are depicted in Figure \ref{fig:optimal_configurations_RSL}.
We note that there exists a region where all the points with high goodness score are concentrated and that there is a sharp transition between the red and the blue region.
This region corresponds to definite values of $\hat{a}_{c} \simeq 0.04$ and $\hat{a}_{l} \simeq 2$,
but there is no definite value for $R$.
In fact the ``optimal'' region is a line, along which $R$ can vary between 1 and 20.
Thus, in the following we choose the maximum value $\hat{R}=20$.

\subsection{Validation of the agent-based model}
\label{sec:valid-agent-based}

To challenge the validity of our agent-based model with the calibrated parameters, we need to reject the Null hypothesis $H_{0}$ that the parameters $(\hat{a}_{c}, \hat{a}_{l},\hat{R})$, are correct. 

\begin{figure}[htbp]
\begin{center}
\includegraphics[width=0.49\textwidth]{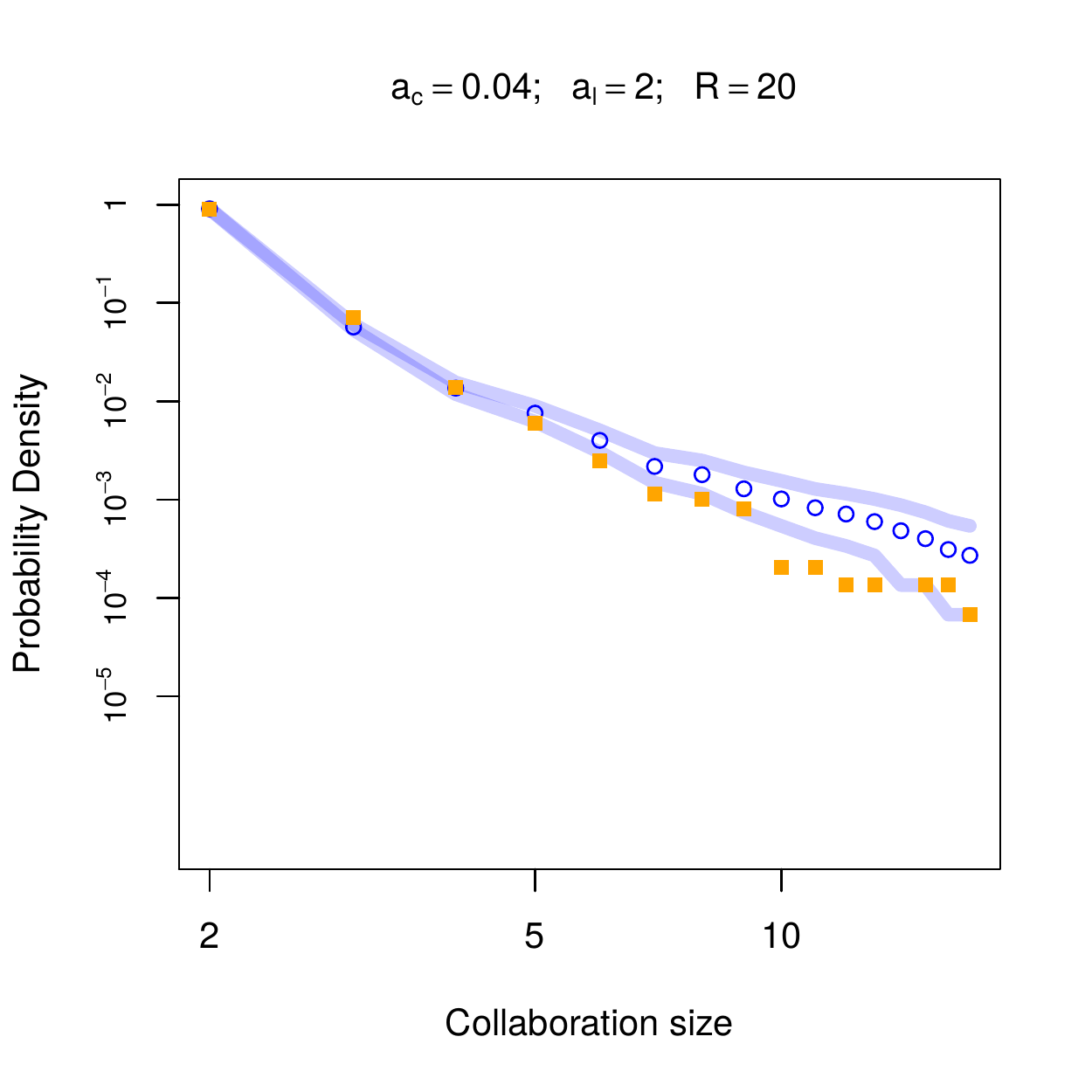}
\end{center}
\vspace{-18pt}
\caption[]{Alliance size distribution of the R\&D network obtained from our agent-based model using the parameters from the maximum likelihood estimation.
  Orange squares represent the empirical size distribution, $p_{s}^{e}(s)$,  blue circles the theoretical size distribution, $p_{s}(s)$ and  blue lines the theoretical 95\% quantiles obtained of the simulations of the multinomial distribution $M$. They define a confidence region for the theoretical size distribution. 
}
\label{fig:optimal_configurations_RD}
\end{figure}

The assumption of our model is that all alliance formations are independent.
This implies that our data has been generated by a multinomial distribution 
${M}\sim Mult\left(N,p_{\tilde{s}}(2),p_{\tilde{s}}(3),\ldots, p_{\tilde{s}}(n)\right)$, where the probabilities $p_{\tilde{s}}(\tilde{s})$ result from our agent-based model.
We use this multinomial distribution to construct a region of $95\%$ probability coverage, which we estimate by sampling $10^6$ times from $M$. 
In Figure~\ref{fig:optimal_configurations_RD} we represent this region as bands around the alliance size distribution $p_{s}(s)$ obtained from the maximum likelihood approach. 
One band corresponds to the $0.025$ quantile, the second one to the $0.975$ quantile for the probabilities $p_{\tilde{s}}(\tilde{s})$.

A comparison with the empirical distribution $p_{s}^{e}(s)$ reveals the very good results.
We see that in most of the empirical values are within the $95\%$ confidence region.
The four data points outside of that region are at least very close to the lower band.
They also represent alliance sizes of only very small probabilities, about $10^{-4}$, and thus cannot be considered as important deviations from the model.
Hence, we conclude that our null hypothesis that the data are generated by the calibrated agent-based model \emph{cannot} be rejected.

\section{Conclusions}
\label{sec:conclusions}

Our goal in this paper is to explain the empirically observed alliance size distribution, Figure~\ref{fig:m_partners}, of R\&D networks by means of an agent-based model that explicitly models the alliance formation process.

Our model builds on \emph{heterogeneous} agents characterized by one individual parameter, their fitness $\eta_{i}$.
The distribution of fitness values is proxied by the empirical activity distribution, Figure~\ref{fig:empirical_fitness_attributes}, as the only model input.
Activity describes how often an agent was engaged in an alliance during the observation period, which is 26 years in our case.
This can be seen as an indication of the agent's attractiveness for other agents to collaborate with, and fitness should be interpreted in the same manner.
We have shown that the fitness distribution obtained this way is right skewed and very broad. 

Further, our agent-based model uses three free parameters that need to be determined in comparison with empirical data.
The calibration process is based on a maximum likelihood estimation that returns those parameter values that match best the target, which is the empirical distribution of alliance sizes.

It is interesting to note that only two of these parameters, the scaling factors $a_{c}$ for the cost of the consortium and $a_{l}$ for the individual threshold to accept an invitation obtain a stable value in the maximum likelihood estimation, whereas the third parameter $R$, the number of rejections to stop forming an alliance does not reach a definite value.
Instead, we observe that equally good likelihoods are obtained for a larger range of $R$ between 1 and 20.
Hence, our model works \emph{without} assuming a specific value of $R$.
In other words, $R$ can vary across time, industrial sectors or even alliances without questioning the validity of our model.


For our model validation we used a high number of rejections, $\hat{R}=20$.
This is definitely realistic for a system such as the global, inter-sectoral R\&D network that we analyze.
Here, firms have to search for their partners among a huge number of potential candidates, making the establishment of an R\&D alliance potentially costly and risky.
We argue that this leads to a very long and cautious selection process, from the side of both the initiator and the invited firm.
Therefore, firms have to be willing to accept a high number of rejections, if they want to gain access to external knowledge and eventually establish R\&D collaborations with other firms.

Regarding the other two parameters, $\hat{a}_{c}=0.04$ and   $\hat{a}_{l}=2$, we note from Equation \eqref{eq:9} that actually their ratio matters, as it determines the range of fitness values $[0,\eta^{\star}]$ for which agents join an alliance. 
The definite value of $\hat{a}_{c}$  should be interpreted as rather large.
I.e. when multiplied with the size of the alliance, the cost in Equation \eqref{eq:1} is rather high in comparison with the benefit of the alliance, which is the sum of the fitness values of the agents.
This has two consequences.
First, it restricts the maximum \emph{size} of an alliance to values below 20.
Second, it restricts the maximum \emph{number} of alliances with sizes larger than 2, because most agents in the system have a rather low fitness and are thus not able to overcome the considerable cost of forming an alliance. 
To illustrate this, an alliance of two agents with median fitness values exhibits a benefit of 0.004 and a cost of 0.04; or an alliance of four agents with median-fitness agents exhibits a benefit  of 0.008 and a cost of 0.12, i.e. almost an order of magnitude larger.

This reflects the intention of our agent-based model.
Agents with high fitness  (typically incumbent firms) are the ones that are most likely to receive an invitation.
At the same time, they will most often refuse the invitation, if an alliance consists of only agents with medium or low fitness nodes (typically mid-size firms or startups).
This leads to the high value of rejections obtained from the maximum likelihood estimation.
But if agents with a high fitness initiate an alliance, agents with medium or low fitness are likely to accept this invitation.
On the other hand, agents with low fitness are not very selective to refuse any invitation because their threshold utility $u_{i}^{\mathrm{thr}}$ is rather low, also as a consequence of the small value of $\hat{a}_{l}$.

The good match of our agent-based model with the empirical observations allows us to draw some conclusions about the formation of real R\&D alliances, for which no data is available. 
As we have seen, alliances are more likely initiated by an incumbent firm of high fitness which directs its interest toward a mid-size company or a startup.
At the same time,
the ``bottleneck'' in establishing new alliances is probably on the initiator's side, which has to take rejections and keep looking for new partners until it finds the right one.




Our agent-based model was developed to reproduce the empirical distribution of alliance \emph{sizes}. 
One could be interested to know whether this model, using the parameter from the maximum likelihood estimation, is also able to reproduce other features of the observed topology of the R\&D network.
This is not the aim of the paper, but we can comment at least on the degree distribution which was analyzed already by \citet{tomasello2014therole}. 
\emph{Degree} refers to the \emph{number} of collaboration \emph{partners} of an agent, not to the number of alliances the agent is involved.
As such, degree is not independent of the size of an alliance, and indeed the empirical degree distribution was also shown to be right skewed and very broad.

However, we argue that the degree distribution \emph{cannot} simply be obtained from our agent-based model because this does not take \emph{degree-degree correlations} into account.
\emph{Assortativity} reflects the tendency of agents with \emph{high} degree to form alliances with other agents with \emph{high} degree, whereas \emph{dissortativity} would indicate that agents with \emph{high} degree have the tendency to form alliances with agents of \emph{low} degree.
Such degree-degree correlations have been detected by \citep{tomasello2013riseandfall} both for sectoral R\&D networks and for the aggregated R\&D network used in this paper.
They play a role in particular for agents with high degree.
Therefore, we can assume that our agent-based model will be able to reproduce the right skewed and broad degree distribution, but becomes increasingly worse in the range of larger degrees.

To conclude, our agent-based model provides  a considerable step forward in identifying the real mechanisms for alliance formation \citep{ahuja2000duality}.
In particular, with the distribution of alliance sizes we are able to reproduce a feature that has received some attention in the existing literature, but never a conclusive explanation.
Our model can be used for stochastic agent-based simulations, it also provides an analytical solution that considerably reduces the computational effort. 
We emphasize that it is rather rare to obtain an analytic description of an agent based model.
Our derivations also apply to cases with different cost functions and are thus quite general.
This should inspire further agent based modelling approaches.

Our agent-based model is fully calibrated and validated against real data from the global interfirm R\&D network.
It shows the emergence of a broad, right-skewed distribution of alliance sizes, taking into account a heterogeneous fitness distribution of agents. 
On the methodological side, our study provides an approach to infer the correct parameter values for the agent-based model, to interpret them and check their consistency with reality.
Like for any agent-based model approach, we cannot conclude that our model is the only one able to explain and reproduce the alliance size distribution.
However, the very good match with reality is a clear sign of plausibility for the set of agent rules that we propose, thus providing us with new insights into the micro dynamics of alliance formation.

\section*{Acknowledgements}
M.V.T. and F.S. acknowledge financial support from the Swiss National Science Foundation, through grant 100014\_126865, ``R\&D Network Life Cycles''.
M.V.T. and F.S. acknowledge financial support from the Seed Project SP-RC 01-15 ``Performance and resilience of collaboration networks'', granted by the ETH Risk Center of ETH Zurich.
The authors thank Ryan Murphy for comments on an early version of this paper.

\bibliographystyle{sg-bibstyle} 


\end{document}